\documentclass[aps,prl,showpacs,amsfonts,amsmath, amssymb,twocolumn,floatfix, superscriptaddress]{revtex4-1}
\usepackage[final]{graphicx}
\usepackage{xcolor}
\usepackage{verbatim}
\usepackage{csquotes}
\usepackage{braket}
\usepackage{mathtools}
\usepackage[colorlinks=true, linkcolor=blue, citecolor=blue, urlcolor=blue]{hyperref}

\definecolor{green}{rgb}{0.2,0.6,0.2}

\newcommand{\vct}[1]{\mathbf{#1}}

\newcommand{\be}{\begin{equation}}
\newcommand{\ee}{\end{equation}}
\allowdisplaybreaks

\begin{document}


\title{Using the fluctuation-dissipation theorem for nonconservative forces}

\author{Kiryl Asheichyk}
\email[]{asheichyk@is.mpg.de}
\affiliation{4th Institute for Theoretical Physics, Universit\"at Stuttgart, Pfaffenwaldring 57, 70569 Stuttgart, Germany}
\affiliation{Max Planck Institute for Intelligent Systems, Heisenbergstrasse 3, 70569 Stuttgart, Germany}
\author{Matthias Kr\"uger}
\email[]{matthias.kruger@uni-goettingen.de}
\affiliation{Institute for Theoretical Physics, Georg-August-Universit\"at G\"ottingen, 37073 G\"ottingen, Germany}

\begin{abstract}
An equilibrium system which is perturbed by an external potential relaxes to a new equilibrium state, a process obeying the fluctuation-dissipation theorem. In contrast, perturbing by nonconservative 
forces yields a nonequilibrium steady state, and the fluctuation-dissipation theorem can in general not be applied. Here we exploit a freedom inherent to linear response theory: Force fields which 
perform work that does not couple statistically to the considered observable can be added without changing the response. Using this freedom, we demonstrate that the fluctuation-dissipation theorem can 
be applied for certain nonconservative forces. We discuss the case of a nonconservative force field linear in particle coordinates, where the mentioned freedom can be formulated in terms of symmetries. 
In particular, for the case of shear, this yields a response formula, which we find advantageous over the known Green-Kubo relation in terms of statistical accuracy.
\end{abstract}


\maketitle



The linear response of a classical equilibrium system to a potential perturbation $ U^{\rm ptb} $ applied for time 
$ t > 0 $ is given by the fluctuation-dissipation theorem (FDT)~\cite{Callen1951, Kubo1966, Marconi2008, Hansen2009},
\begin{equation}
{\langle A(t) \rangle}^{\rm ptb} - \langle A \rangle = - \frac{1}{k_{\rm B}T}\left[\langle AU^{\rm ptb}\rangle - \langle A(t) U^{\rm ptb}(0) \rangle\right],
\label{eq:FDT_pp}
\end{equation}
where $ A $ is an observable of interest, $ k_{\rm B} $ is Boltzmann's constant, $ T $ is  temperature, and $ {\langle \cdots \rangle}^{\rm ptb} $ and 
$ \langle \cdots \rangle $ indicate averages in the perturbed and equilibrium system, respectively. The stationary limit of
formula~\eqref{eq:FDT_pp} can be derived from the equilibrium distribution the system relaxes to.

In contrast, a nonpotential perturbation drives the system to a nonequilibrium steady state. The corresponding (nonequilibrium) distribution is typically unknown \cite{Seifert2012}, and the linear 
responses to these types of perturbations yield forms fundamentally different from Eq.~\eqref{eq:FDT_pp}. One hence applies other methods in this case, as equations for a probability 
distribution~\cite{Hansen2009, Gardiner2009, Kubo1991, Risken1996}, path integral techniques~\cite{Altland2010, Martin1973, Janssen1976, DeDominicis1978, Onsanger1953, Machlup1953} or Malliavin 
calculus~\cite{Nualart2006, Bell2006, Warren2012}. Compared to Eq.~\eqref{eq:FDT_pp}, the resulting response relations show less universality and typically contain time integration through the transient 
dynamics of the system (see Refs.~\cite{Marconi2008, Warren2012, Fuchs2005, Chong2009, Suzuki2013, Sharma2016} for specific examples).

In this work, we exploit a simple method for computing the linear response to a nonpotential perturbation via FDT for certain observables, using the freedom of adding forces whose work does not couple 
to the considered observable. Note that a similar freedom has been discussed in Ref.~\cite{Seifert2010}. We illustrate that, for a force perturbation linear in coordinates, the mentioned freedom 
can be formulated in terms of symmetries. We study in detail the case of a Brownian system perturbed by simple shear flow, finding a response formula [Eq.~\eqref{eq:RK} below], which is an 
alternative to the classical Green-Kubo relation [Eq.~\eqref{eq:GK} below]. Using numerical simulations, the formula is found to have a lower variance for all cases studied, making it advantageous 
in terms of statistics.

Consider a classical system of $ N $ interacting particles, subject to external potentials and coupled to a heat bath at temperature $ T $, in thermal equilibrium at time $ t = 0 $. For time $ t > 0 $, 
the system is perturbed by nonconservative forces $ \{\vct{F}^{\rm ptb}_i\} $, with $\vct{F}^{\rm ptb}_i$ acting on particle $i$ at position $ \vct{r}_i $. Because the equilibrium state is 
time symmetric, the linear response of $A$ is related to the work done on the system~\cite{Marconi2008, Kurchan1998}, 
\begin{equation}
\langle A(t) \rangle^{\rm ptb} - \langle A \rangle = \frac{1}{k_{\rm B}T}\int_0^t dt' \left\langle A(t)\sum_{i=1}^N\vct{F}^{\rm ptb}_i(t')\cdot\dot{\vct{r}}_i(t')\right\rangle.
\label{eq:FDT_general}
\end{equation}
If $\vct{F}^{\rm ptb}_i $ is a conservative force, i.e., $\vct{F}^{\rm ptb}_i=-\nabla_i U^{\rm ptb}$, the work 
$\int_0^{t}dt' \sum_{i=1}^N\vct{F}^{\rm ptb}_i(t')\cdot\dot{\vct{r}}_i(t') = U^{\rm ptb}(0) - U^{\rm ptb}(t) $ depends only 
on the states, and Eq.~\eqref{eq:FDT_pp} follows from Eq.~\eqref{eq:FDT_general}. 

Notably, response relation~\eqref{eq:FDT_general} displays a freedom in $\vct{F}^{\rm ptb}_i$ when computing the perturbed $A$: It allows adding perturbation forces $ \vct{G}^{\rm ptb}_i $ whose work 
does not couple to the observable $A$, i.e.,
\begin{equation}
\int_0^t dt' \left\langle A(t)\sum_{i=1}^N\vct{G}^{\rm ptb}_i(t')\cdot\dot{\vct{r}}_i(t') \right\rangle = 0,
\label{eq:freedom}
\end{equation}
without changing the response of $A$. Thus, if a force $ {\vct{G}}^{\rm ptb}_i $ obeying Eq.~\eqref{eq:freedom} exists such that adding the two forces results in a potential $U^{\rm ptb} $,
\begin{equation}
\vct{F}^{\rm ptb}_i + \vct{G}^{\rm ptb}_i = -\nabla_iU^{\rm ptb},
\label{eq:FplusG}
\end{equation}
then, according to Eq.~\eqref{eq:FDT_general}, the response of $A$ to the nonconservative force $ \vct{F}^{\rm ptb}_i $ is equivalent to the response to the potential $ U^{\rm ptb} $ and 
given by formula~\eqref{eq:FDT_pp}. Exploring this possibility of restoring an FDT is the content of this paper.

\begin{figure}[!t]
\includegraphics[width=1.0\linewidth]{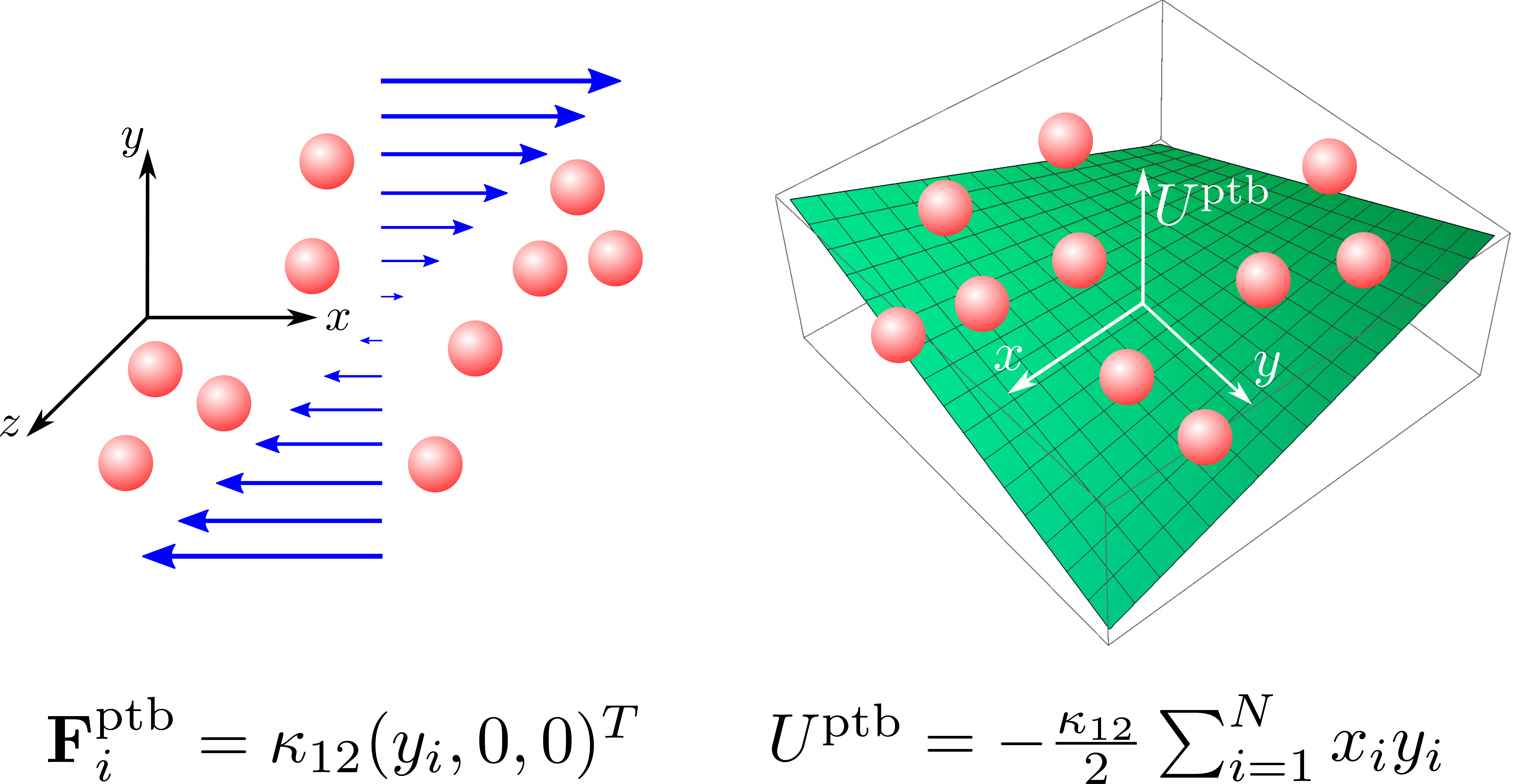}
\caption{\label{fig:method}Illustration of the concept for the case of simple shear, $ \vct{F}^{\rm ptb}_i = \kappa_{12}(y_i,0,0)^T $. Superposition of $ \vct{F}^{\rm ptb}_i$ and 
$ \vct{G}^{\rm ptb}_i $ given by Eq.~\eqref{eq:G_flow} results in the gradient of the potential 
$ U^{\rm ptb}  = -\frac{\kappa_{12}}{2}\sum_{i=1}^Nx_iy_i  $. Note that this corresponds to superposition of the shear field with its image mirrored at the plane $ x = y $. Given the 
symmetries detailed in the main text, the linear responses to $ \vct{F}^{\rm ptb}_i $ and $ U^{\rm ptb} $ are identical.}
\end{figure}

We investigate the specific case of a force field linear in $\vct{r}_i$,
\begin{equation}
\vct{F}^{\rm ptb}_i = \boldsymbol{\kappa}\cdot\vct{r}_i,
\label{eq:flow_ptb}
\end{equation}
with the tensor $ \boldsymbol{\kappa} $ independent of particle positions. If $ \boldsymbol{\kappa} $ is symmetric, $\vct{F}^{\rm ptb}_i$ derives from a generalized harmonic potential. The case of 
interest is that $ \boldsymbol{\kappa} $ is  not symmetric, such that $\vct{F}^{\rm ptb}_i$ of Eq.~\eqref{eq:flow_ptb} is not conservative. One natural way of exploring the above-mentioned freedom is by 
using  the transpose of $ \boldsymbol{\kappa} $, i.e., it is promising to use
\begin{equation}
\vct{G}^{\rm ptb}_i = \frac{1}{2}\left(\boldsymbol{\kappa}^T - \boldsymbol{\kappa}\right)\cdot\vct{r}_i.
\label{eq:G_flow}
\end{equation}
The sum of $\vct{F}^{\rm ptb}_i$ and $ \vct{G}^{\rm ptb}_i$ is then immediately found,
\begin{equation}
\vct{F}^{\rm ptb}_i+ \vct{G}^{\rm ptb}_i = \frac{1}{2}\left(\boldsymbol{\kappa} + \boldsymbol{\kappa}^T\right)\cdot\vct{r}_i = -\nabla_i  U^{\rm ptb}(\{\vct{r}_i\}),
\label{eq:flow_ptb_potential}
\end{equation}
where the potential is identified as 
\begin{equation}
U^{\rm ptb}(\{\vct{r}_i\}) = -\frac{1}{4}\sum_{i=1}^N \vct{r}_i \cdot \left(\boldsymbol{\kappa} + \boldsymbol{\kappa}^T\right)\cdot\vct{r}_i.
\label{eq:flow_ptb_potential_explicit}
\end{equation}

How to satisfy Eq.~\eqref{eq:freedom}? Many cases that do so can be identified on the basis of symmetries, as we demonstrate by regarding 
$ \boldsymbol{\kappa} = \kappa_{12}\hat{\vct{x}}\otimes \hat{\vct{y}}$ (with $\hat{\vct{x}}$, $\hat{\vct{y}}$, and $ \otimes $ denoting unit vectors and the tensor product, respectively), i.e., shear 
forces (see Fig.~\ref{fig:method} for an illustration)~\footnote[1]{Due to the superposition principle for linear responses and the fact that a general linear force can be decomposed into a potential 
part and a sum of shear forces in different directions, the presented example of $\boldsymbol{\kappa} = \kappa_{12}\hat{\vct{x}}\otimes \hat{\vct{y}}$ can be used to obtain an FDT for any 
$\boldsymbol{\kappa} $.}. From Eq.~\eqref{eq:flow_ptb_potential_explicit}, the corresponding potential reads 
\begin{equation}
U^{\rm ptb} = -\frac{\kappa_{12}}{2}\sum_{i=1}^Nx_iy_i,
\label{eq:shear_potential}
\end{equation}
being a potential with one stable and one unstable direction in the $xy$ plane (see Fig.~\ref{fig:method}). One direct way of fulfilling Eq.~\eqref{eq:freedom} is restricting to systems and 
observables which are symmetric under interchange of the $ x $ and $ y $ coordinates. These are systems for which interaction and external potentials remain the same under 
interchange $ \{x_i\} \leftrightarrow \{y_i\} $, and observables which remain the same under interchange $ \{x_i\} \leftrightarrow \{y_i\} $ and $ \{v_{ix}\} \leftrightarrow \{v_{iy}\} $ (where 
$ v_{ix} $ denotes the $ x $ component of the velocity of particle $i$). Then condition~\eqref{eq:freedom} is fulfilled by symmetry~\footnote[2]{This can be understood from the fact that 
$ \vct{G}^{\rm ptb}_i $, given by Eq.~\eqref{eq:G_flow}, is a difference between shear forces in the $ y $ and $ x $ directions, whose works averaged with $ A $ are identical in the mentioned case of 
$ xy $ symmetry.}.  
For example, spherically symmetric potentials and observables like $ A = \sum_{i=1}^Nx_iy_i $, $ A = \sum_{i=1}^Nv_{ix}v_{iy} $, or the $ xy $ component of the stress tensor~\cite{Hansen2009} 
comprise these symmetries. 
Substituting Eq.~\eqref{eq:shear_potential} into Eq.~\eqref{eq:FDT_pp}, we find that, for these cases, the linear response to shear forcing is given by
\begin{align}
\notag & {\langle A(t) \rangle}^{\rm ptb} - {\langle A \rangle} \\
& = \frac{\kappa_{12}}{2k_{\rm B}T}\left[\left\langle A\sum_{i=1}^Nx_iy_i \right\rangle - \left\langle A(t)\sum_{i=1}^Nx_i(0)y_i(0) \right\rangle\right].
\label{eq:RK0}
\end{align}
Formula~\eqref{eq:RK0} thus provides the response to a nonconservative force (shear force) via FDT.

Many models treat forces and external flow driving in an identical manner. Regarding Brownian particles with mobility $ \mu $ [see Eq.~\eqref{eq:LE1} below], an external flow velocity field 
$\vct{V}(\vct{r})$ gives rise to a force $\vct{F}^{\rm ptb}_i = \frac{\vct{V}(\vct{r}_i)}{\mu}$ \cite{Dhont1996}, neglecting hydrodynamic interactions. The mentioned shear \textit{forces} then translate 
to shear \textit{flow} by identifying $\kappa_{12} = \frac{\dot\gamma}{\mu}$ (with shear rate $\dot\gamma$), and, under the above symmetries, we obtain for the shear perturbed 
${\langle A(t) \rangle}^{(\dot\gamma)}$,
\begin{align}
\notag & {\langle A(t) \rangle}^{(\dot\gamma)} - {\langle A \rangle} \\
& = \frac{\dot\gamma}{2k_{\rm B}T\mu}\left[\left\langle A\sum_{i=1}^Nx_iy_i \right\rangle - \left\langle A(t)\sum_{i=1}^Nx_i(0)y_i(0) \right\rangle\right].
\label{eq:RK}
\end{align}
Response relations~\eqref{eq:RK} and \eqref{eq:RK0} are our main results. We note that Eq.~\eqref{eq:RK} has been derived in 
Ref.~\cite{Warren2012} for a single overdamped Brownian particle. A prestage version of Eq.~\eqref{eq:RK} has been given in Ref.~\cite{Asheichyk2019}, and we discuss the
relation at the end of this paper. Equation~\eqref{eq:RK} is an alternative to the classical Green-Kubo relation for shear which, for the case of 
overdamped Brownian particles, reads as [see Refs.~\cite{Kubo1966, Marconi2008, Hansen2009, Kubo1991, Fuchs2005, Green1954, Kubo1957} for various Green-Kubo relations, and Ref.~\cite{Fuchs2005} for 
formula~\eqref{eq:GK} in particular]
\begin{equation}
{\langle A(t) \rangle}^{(\dot{\gamma})} - \langle A \rangle = \frac{\dot{\gamma}}{k_{\rm B}T}\int_0^tdt'\left\langle A(t')\sigma_{xy}(0)\right\rangle,
\label{eq:GK}
\end{equation}
where $ \sigma_{xy} $ is the $ xy $ component of the stress tensor defined as $ \sigma_{xy} = -\sum_{i=1}^N\left(F_{ix}^{\rm int}+F_{ix}^{\rm ext}\right)y_i $, with $ F_{ix}^{\rm int} $ and 
$ F_{ix}^{\rm ext} $ being interaction and external forces, respectively, acting on particle $ i $ in direction $ x $.
One advantage of Eq.~\eqref{eq:RK} over Eq.~\eqref{eq:GK} is the absence of a time integral. Another is that forces do not have to be measured.  

Seeking a numerical example, we turn to interacting overdamped Brownian particles in two space dimensions, following Langevin dynamics~\cite{Marconi2008, Kurchan1998},  
\begin{equation}
\frac{\dot{\vct{r}}_i}{\mu} = \boldsymbol{\kappa}\cdot\vct{r}_i  +  \vct{F}^{\rm int}_i +  \vct{F}^{\rm ext}_i +  \vct{f}_i,
\label{eq:LE1}
\end{equation}
where $ \mu\boldsymbol{\kappa}\cdot\vct{r}_i $ is the shear velocity, imposed at $ t > 0 $, with the shear-rate tensor $ \mu\boldsymbol{\kappa} = \dot{\gamma}\hat{\vct{x}}\otimes \hat{\vct{y}} $. 
\begin{equation}
\vct{F}^{\rm int}_i = -\nabla_i \frac{\Gamma}{2}\sum_{i=1}^N\sum_{j=1(j\neq i)}^N\frac{1}{r_{ij}} e^{-\frac{r_{ij}}{r_c}}
\label{eq:Fint}
\end{equation}
are interaction forces, chosen to arise from a screened Coulomb potential, with interparticle distance $ r_{ij}\equiv|\vct{r}_i-\vct{r}_j|$, coupling 
strength $\Gamma$, and interaction range $r_c$. The external force follows from a harmonic potential,
\begin{equation}
\vct{F}^{\rm ext}_i =-\nabla_i \frac{k}{2}\sum_{i=1}^N|\vct{r}_i|^2,
\label{eq:Fext}
\end{equation}
with spring constant $k$.
$ \vct{f}_i $ is a Gaussian white noise,
\begin{equation}
\langle \vct{f}_i(t) \rangle = 0, \ \ \ \langle \vct{f}_i(t) \otimes \vct{f}_j(t') \rangle = \frac{2k_{\rm B}T}{\mu}\mathbb{I}\delta_{ij}\delta(t-t'),
\label{eq:RFProp}
\end{equation}
where $ \mathbb{I} $ is the identity matrix. We set $ k_{\rm B}T = r_c = \mu = 1 $, and $ \Gamma = 25 $. $ N $, $ k $, $ \dot{\gamma} $, and the number of 
independent noise realizations $ C $ for performing averages are varied between measurements. The dynamics is simulated using the Euler method. We choose $ A = \sum_{i=1}^Nx_iy_i $, 
which is the lowest nontrivial moment of the particle distribution. Since the system and $ A $ are $ xy $ symmetric, condition~\eqref{eq:freedom} is fulfilled and formula~\eqref{eq:RK} is valid. 

\begin{figure}[!t]
\includegraphics[width=1.0\linewidth]{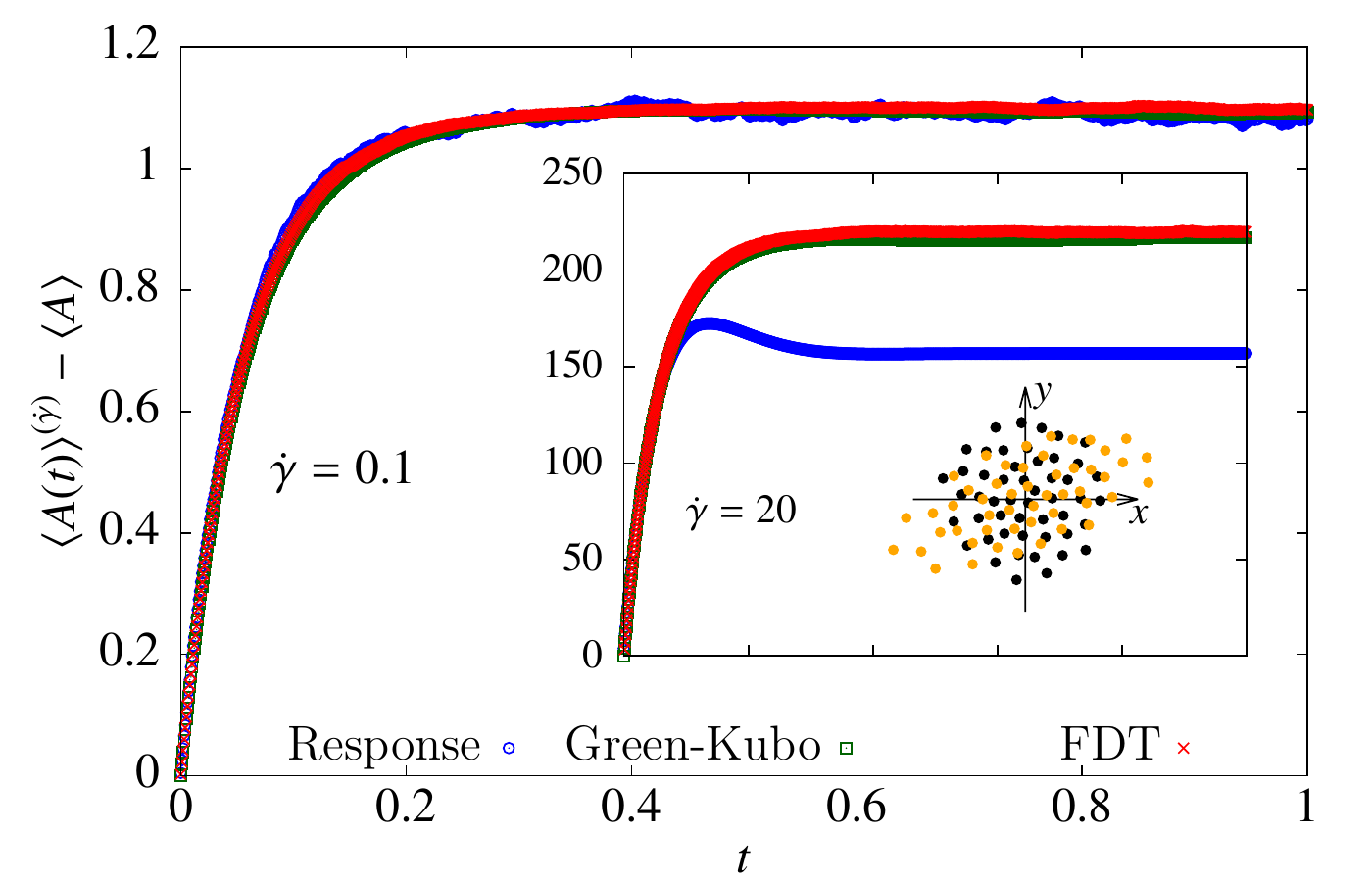}
\caption{\label{fig:testRK}Response to shear flow for $ A = \sum_{i=1}^Nx_iy_i $ of a two-dimensional system of interacting Brownian particles confined in a harmonic trap. The main plot shows 
the linear (small $ \dot{\gamma} $) response computed by shearing (\enquote{Response}), using the Green-Kubo formula~\eqref{eq:GK}, and using Eq.~\eqref{eq:RK} (\enquote{FDT}). The 
inset plot shows a nonlinear (large $ \dot{\gamma} $) response with the corresponding simulation snapshots demonstrating the effect of shear on the morphology of the cluster (black particles are in 
equilibrium, while orange particles are sheared). Parameters: $ N = 50 $, $ k = 10 $, and $ C = 4 \times 10^5 $.}
\end{figure}

We compute $ {\langle A(t) \rangle}^{(\dot{\gamma})} - {\langle A \rangle} $ via three different routes: by (i) applying finite shear, (ii) using equilibrium correlations according to the 
Green-Kubo formula~\eqref{eq:GK}, and (iii) using equilibrium correlations according to Eq.~\eqref{eq:RK} (labeled \enquote{FDT} in the figures). Figure~\ref{fig:testRK} compares 
these  as a function of time $t$ after start of shear. For small shear rate (main plot), all methods agree, thereby confirming formula~\eqref{eq:RK}. For large shear rate (inset plot), the deviation 
from the linear response is evident, also regarding the form of the response curve, which shows a characteristic \enquote{overshoot,} i.e., a nonmonotonic behavior as a function of time, which has also 
been observed in sheared bulk systems~\cite{Amann2013}. Snapshots for equilibrium (black particles) and sheared (orange particles) systems illustrate the change of shape 
of the cluster from circular to ellipsoidal: $ \langle A \rangle = 0 $, but $ {\langle A(t) \rangle}^{(\dot{\gamma})} \geq 0 $.

\begin{figure*}[!t]
\begin{tabular}{cc}
\includegraphics[width=0.49\linewidth]{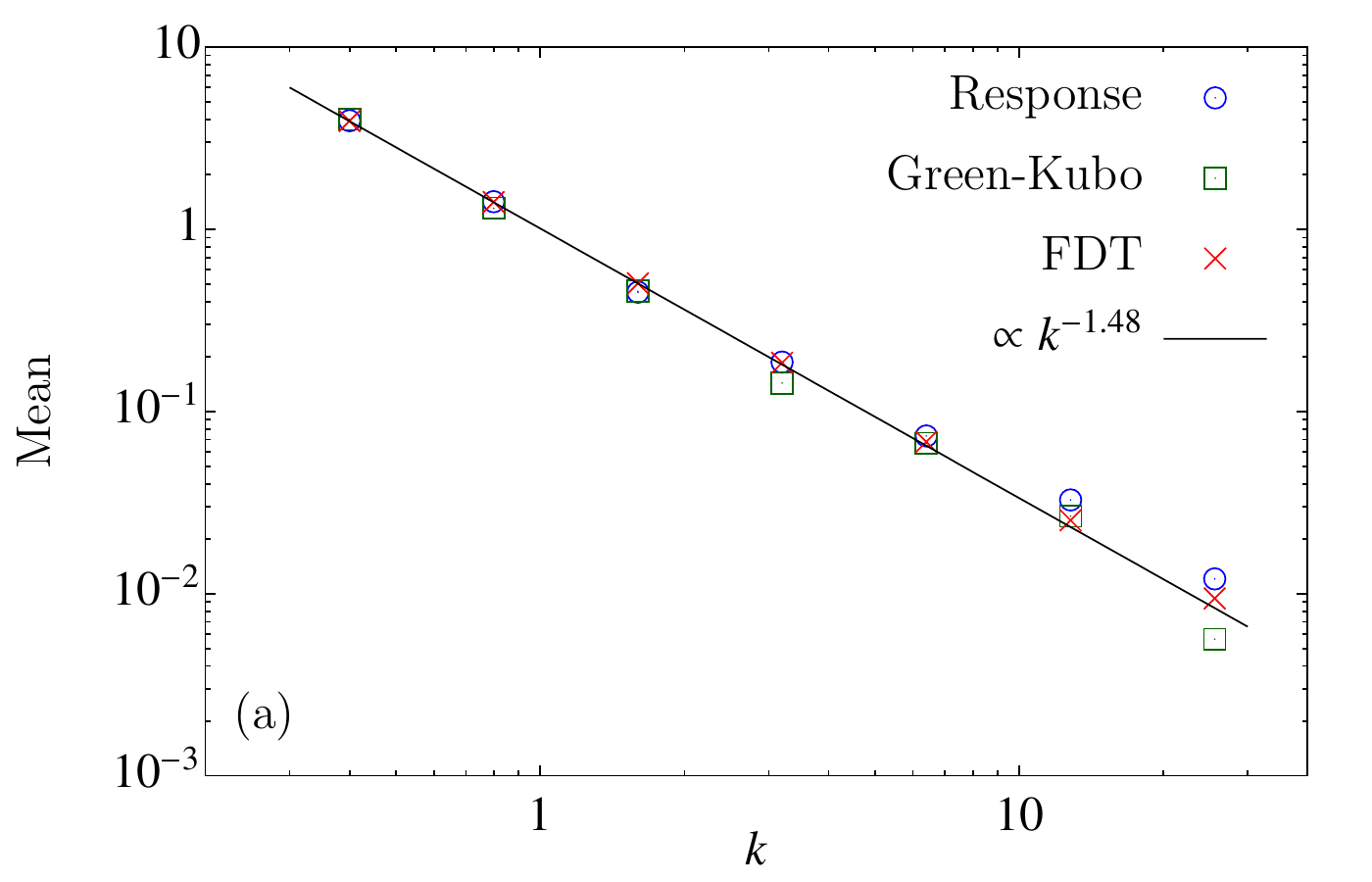}
&
\includegraphics[width=0.49\linewidth]{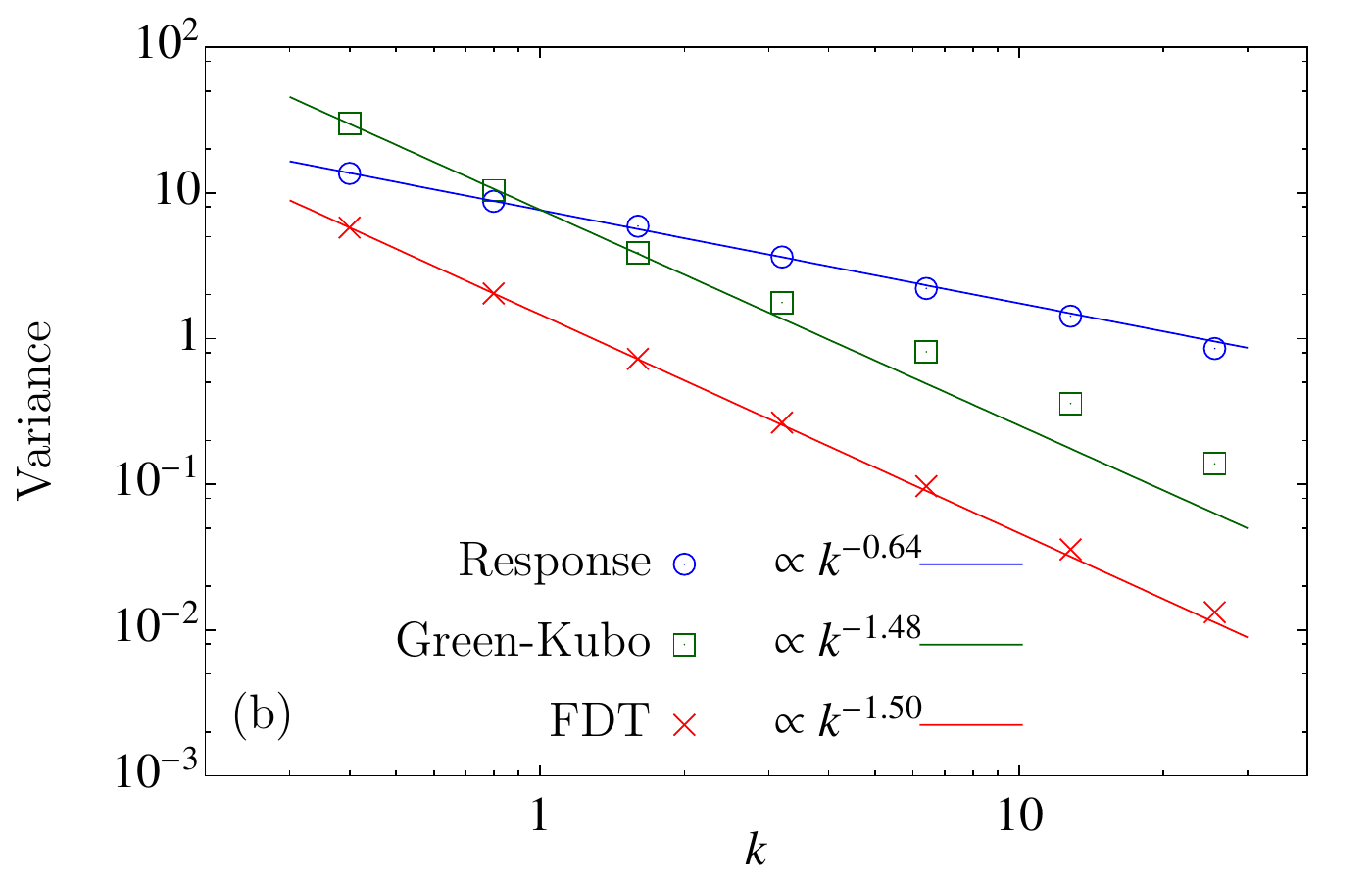}\\
\includegraphics[width=0.49\linewidth]{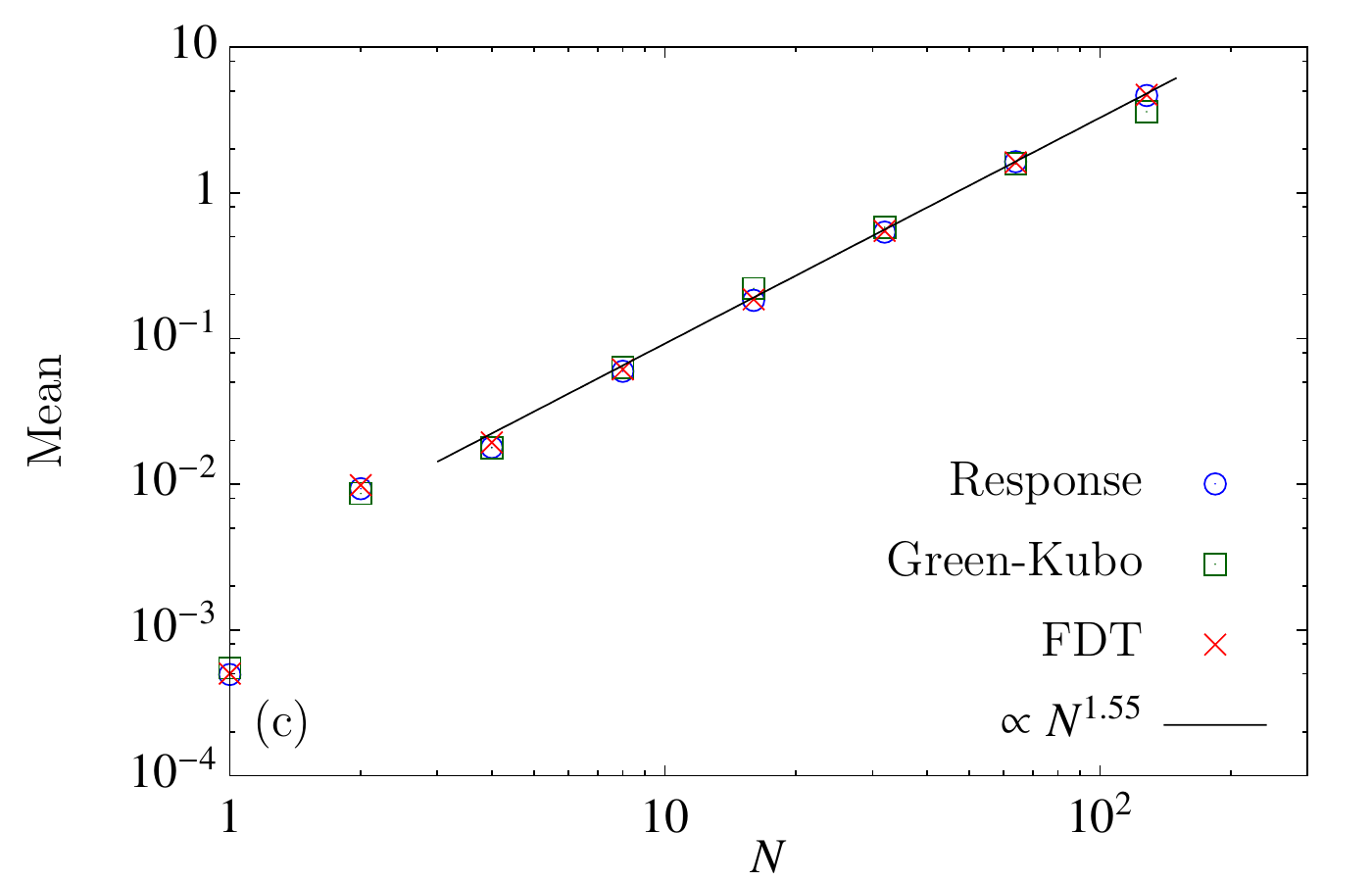}
&
\includegraphics[width=0.49\linewidth]{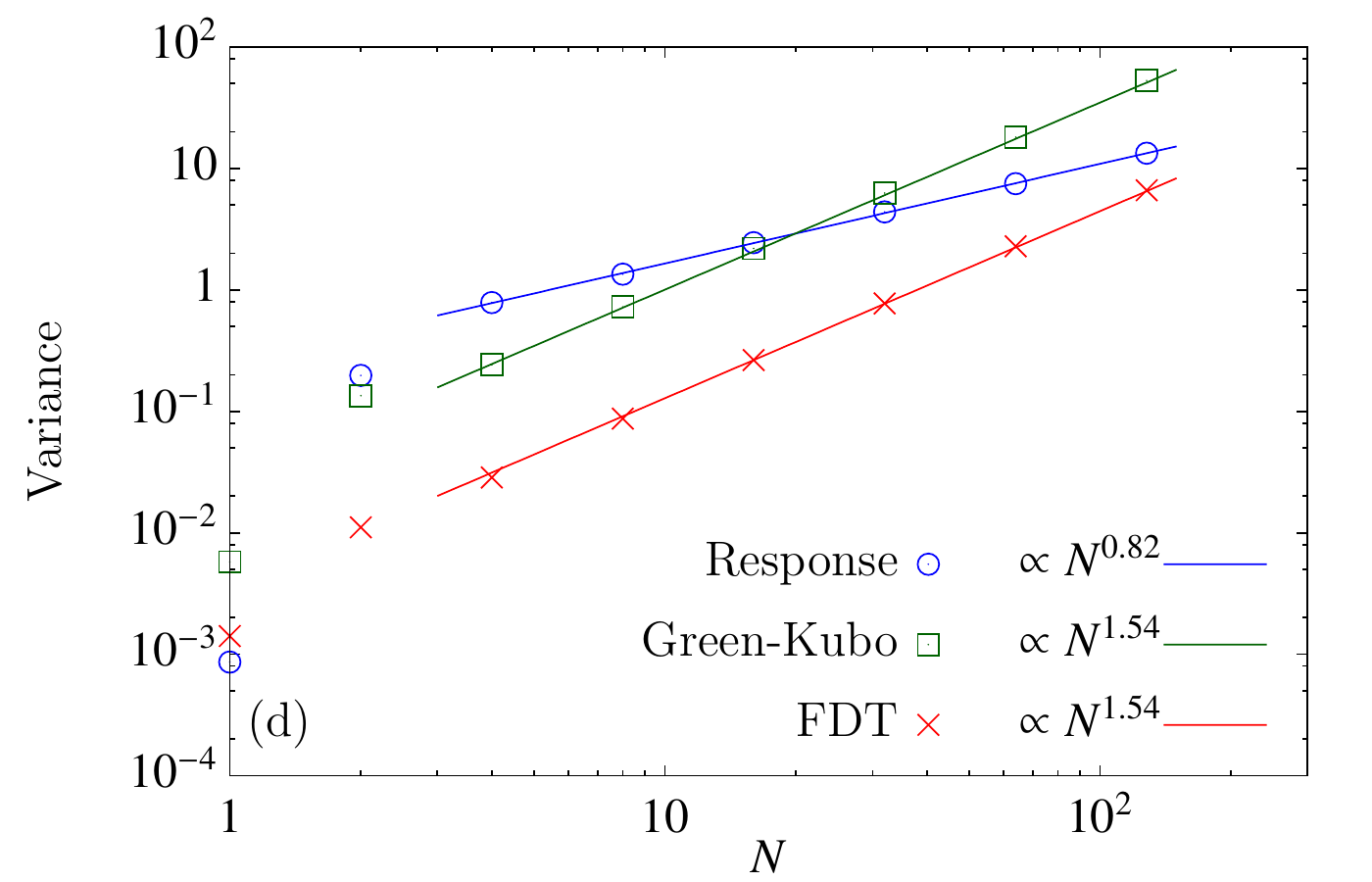}
\end{tabular}
\caption{\label{fig:Var}Dependence of the stationary linear response [(a), (c)] and its variance [(b), (d)] on the confinement strength $ k $ [(a), (b)] and the number of particles $ N $ [(c), (d)] 
obtained in the sheared system (\enquote{Response}), using the Green-Kubo formula~\eqref{eq:GK}, and using Eq.~\eqref{eq:RK} (\enquote{FDT}). Straight lines correspond to power-law fits. Parameters: 
$ N = 10 $ and $ \dot{\gamma} = 0.04 $ for panels (a) and (b); $ k = 10 $ and $ \dot{\gamma} = 0.1 $ for panels (c) and (d).}
\end{figure*}

Panels (a) and (c) of Fig.~\ref{fig:Var} show the dependence on the confinement strength $k$ and the number of particles $N$ of the steady-state response, again confirming agreement between the three 
methods. From fits to the data, the response follows the scaling $ \propto k^{-1.48} $ (compared to $ \propto~k^{-2} $, obtained analytically for $ N = 1 $) and $ \propto N^{1.55} $ (for $ N \agt 4 $).

Panels (b) and (d) of Fig.~\ref{fig:Var} show the corresponding variance, related to the statistical error of a single measurement using the different methods~\footnote[3]{The variance is 
$ \sigma = \sqrt{\langle B^2 \rangle - {\langle B \rangle}^2} $, where $ B = A(\dot{\gamma}) - A(\dot{\gamma}=0) $ [here, noise realizations are chosen to be the same for $ A(\dot{\gamma}) $ and 
$ A(\dot{\gamma}=0) $],  $ B = \frac{\dot\gamma}{2k_{\rm B}T\mu} A\sum_{i=1}^Nx_iy_i $, and $ B = \frac{\dot{\gamma}}{k_{\rm B}T}\int_0^{\tau} dt' A(t')\sigma_{xy}(0) $ for the three methods, 
respectively. $ \tau $ is the time when the steady state of the corresponding mean, {$ \langle B \rangle $}, is reached. We note that the variance for the Green-Kubo relation is not well defined, because 
it does not converge to a stationary value as a function of time.}. It shows a notable difference between the methods following scaling behaviors 
of $ \propto k^{-0.64}N^{0.82} $, $ \propto k^{-1.48}N^{1.54} $, and $ \propto k^{-1.50}N^{1.54} $, respectively. The Green-Kubo relation and formula~\eqref{eq:RK} scale similarly, but the latter has a 
notably lower variance.

\begin{table}[!b]
\caption{\label{table:scaling}Scaling behaviors of the relative variance for the three different computational methods (extracted from Fig.~\ref{fig:Var}).}
\begin{ruledtabular}
\begin{tabular}{c|ccc}
Method & Response & Green-Kubo & FDT\\
Power for $ k $ & $0.84$ & $0$ & $-0.02$ \\
Power for $ N $ & $-0.73$ & $-0.01$ & $-0.01$
\end{tabular}
\end{ruledtabular}
\end{table} 

Table~\ref{table:scaling} compares scaling behaviors of the relative variance (variance divided by the mean) for the three methods. The relative variance of the directly measured response grows with 
$ k $ and decreases with $ N $. For the Green-Kubo relation~\eqref{eq:GK} and for 
Eq.~\eqref{eq:RK}, the relative variance hardly depends on $ k $ and $ N $, indicating that the statistical efficiency of Eqs.~\eqref{eq:GK} and \eqref{eq:RK} is invariant with respect to changes of the 
effective system size and density, highlighting an interesting property of the linear response approach. For the set of parameters we used in our simulations, Eq.~\eqref{eq:RK} has the lowest variance. 
Comparing it to the Green-Kubo relation~\eqref{eq:GK}, it thus needs a much smaller number of independent runs (roughly a factor of $100$ here, estimated from the variance and the central limit theorem), 
which, additionally to the mentioned absence of integration, is advantageous.

Finally, the Langevin equation \eqref{eq:LE1} allows one to give more insights into the nature of Eqs.~\eqref{eq:RK} and \eqref{eq:GK}. Expanding the corresponding path action in shear rate 
$\dot\gamma$ yields for the linear response~\cite{Asheichyk2019}
\begin{align}
\notag {\langle A(t) \rangle}^{(\dot{\gamma})} - \langle A \rangle & = \frac{\dot{\gamma}}{2k_{\rm B}T\mu}\int_0^tdt'\left\langle A(t)\sum_{i=1}^N \dot{x}_i(t')y_i(t')\right\rangle\\
& +\frac{\dot{\gamma}}{2k_{\rm B}T}\int_0^tdt'\left\langle A(t') \sigma_{xy}(0)\right\rangle.
\label{eq:RKGK}
\end{align}
The second term on the right-hand side of Eq.~\eqref{eq:RKGK}, containing the stress tensor, stems from the time-symmetric part of the expanded action, and yields (the half of) Eq.~\eqref{eq:GK}. The 
term containing $\dot{x}_iy_i$ is time antisymmetric, and yields, after adding the transpose shear field, Eq.~\eqref{eq:RK}. Because the equilibrium state is time symmetric, the two terms in 
Eq.~\eqref{eq:RKGK} are identical~\cite{Baiesi2009}. This discussion finally highlights another advantage of FDT and Eq.~\eqref{eq:RK0}: The form of Eq.~\eqref{eq:RK0}, being based on the 
time-antisymmetric part, related to the above-mentioned work, is system independent, while Eq.~\eqref{eq:GK} takes different forms in different systems~\cite{Basu2015}.

Using the freedom of adding forces whose work does not couple to the considered observable, we found that the linear response to nonconservative forces can be computed from FDT. Compared to 
standard approaches, application of this concept typically involves simpler quantities to be measured [e.g., positions in Eq.~\eqref{eq:RK} versus forces in Eq.~\eqref{eq:GK}], does not require time 
integrals, and necessitates a smaller number of independent measurements. This is expected to make it advantageous for both simulations as well as experiments. Future work may address bulk systems via 
Eq.~\eqref{eq:RK} in order to obtain the shear viscosity and to connect to \enquote{Einstein relations} for such 
viscosity~\cite{Hansen2009, Helfand1960, McQuarrie1976, Chialvo1991, Allen1994, Alder1970, Fritschi2018}.

We greatly acknowledge collaboration with Christian~M.~Rohwer and Alexandre P. Solon during the initial stage of this project. We also thank Rosalind~J.~Allen, 
Thomas Voigtmann, and Patrick~B.~Warren for valuable discussions, and thank S. Dietrich and Alexandre~P.~Solon for comments on the manuscript. K.A. is supported by Studienstiftung des deutschen Volkes 
and the Physics Department of the University of Stuttgart. K.A. also acknowledges support by S. Dietrich.



%


\end{document}